\documentclass[5p,10pt,twocolumn]{elsarticle}
\usepackage{amssymb,natbib}
\usepackage{graphicx,subfigure,amsmath,color}
\usepackage{epsfig}
\usepackage{multirow}
\usepackage{rotating}
\journal{Physics Letters B}
\begin{document}
\begin{frontmatter}
\title{Harmonic chiral vibration in triaxial nuclei}
\author[mymainaddress,mymainaddress1]{R. Budaca\corref{mycorrespondingauthor}}
\cortext[mycorrespondingauthor]{Corresponding author}
\ead{rbudaca@theory.nipne.ro; budacaradu@gmail.com}
\author[mymainaddress,address1]{A. I. Budaca}
\address[mymainaddress]{"Horia Hulubei" National Institute for R\&D in Physics and Nuclear Engineering, Str. Reactorului 30, RO-077125, POB-MG6 Bucharest-M\v{a}gurele, Romania}
\address[mymainaddress1]{Academy of Romanian Scientists, 54 Splaiul Independen\c{t}ei, RO-050094, Bucharest, Romania}
\address[address1]{International Centre for Advanced Research and Training in Physics (CIFRA), Str. Atomi\c{s}tilor 409, RO-077125 Bucharest-M\v{a}gurele, Romania}
\begin{abstract}
The low spin states of chiral partner bands are described by means of harmonic oscillations of the total angular momentum vector. The validity of the adopted approximation is established in terms of triaxiality, quasiparticle alignments and the maximal total angular momentum. The spectral properties of the model are described in detail with experimental realizations presented for nuclei with large resultant quasiparticle alignments. The model provides a simple phenomenological way for the determination of the triaxial deformation from the experimental data and easily discernable observable signatures.
\end{abstract}
\end{frontmatter}

When combined with single-particle degrees of freedom, the complex collective rotational behavior of triaxial nuclei can give rise to systems with intrinsic chirality. Such a concept was first proposed in Ref. \cite{FrauMeng} for odd-odd nuclei, having as necessary ingredients a triaxial core and two unpaired quasiparticles from opposite sides of their respective Fermi surfaces. The mean field orbits favored by the holes and particles can generate a trihedral configuration in conjunction with the rotation of the collective core. Consequently, the associated spin vectors can be arranged into a left- or a right-handed configuration. Restoring the chiral symmetry in the laboratory frame of reference should generate a pair of degenerate rotational bands of opposite chirality. The experimental realization of this phenomenon \cite{Petrache} was followed by a widespread identification of chiral bands in many nuclei and its generalization involving more complex structures. On the other hand, its theoretical description evolved based on various models \cite{RadProg,Scripta1,Scripta2,Buletin,PLB1,PLB2}, providing complementing phenomenological insights. The degeneracy of the observed partner bands is however a rare occurrence and limited to short ranges of high spin states. The lack of degeneracy is understood as a soft chiral symmetry breaking associated with a chiral vibration regime at low spins. In this study one will show that this dynamical behaviour can be described by a harmonic oscillation of the total angular momentum vector between chiral configurations.

We start with the rotor Hamiltonian of the particle rotor model (PRM) \cite{BM}
\begin{equation}
\hat{H}_{R}=\sum_{k=1,2,3}A_{k}(\hat{I}_{k}-\hat{j}_{k}-\hat{j'}_{k})^{2},
\end{equation}
expressed in terms of the total angular momentum $\vec{I}$ and two quasiparticle spins $\vec{j}$ and $\vec{j}'$. The inertial parameters $A_{k}=1/(2\mathcal{J}_{k})$ are correlated with the  hydrodynamic moments of inertia
\begin{equation}
\mathcal{J}_{k}=\frac{4}{3}\mathcal{J}_{0}\sin^{2}{\left(\gamma-\frac{2}{3}k\pi\right)},
\label{inert}
\end{equation}
parametrized by the triaxial deformation $\gamma$.

Consider now that the two quasiparticle spins are rigidly aligned along the axes 1 and 2:
\begin{equation}
\hat{j}_{3}=\hat{j'}_{3}=0,\,\,\hat{j}_{1}=j,\,\,\hat{j'}_{2}=j',\,\,\hat{j}_{2}=\hat{j'}_{1}=0.
\end{equation}
With this choice, the chiral geometry is realized by working in the interval $\gamma\in(60^{\circ},120^{\circ})$ of the triaxial deformation. In this way, the maximal moment of inertia will be along the third axis designating the core's rotation. On the other hand, the maximal and minimal extension of the nuclear surface will be distributed along the first and the second body-fixed axes. In conclusion, the $k=1,2,3$ axes are identified as long, short and medium, while the alignment preferences ascribes a hole and particle nature for the quasiparticle spins $\vec{j}$ and $\vec{j}'$. Leaving aside the renaming of the principal axes, the present range of the triaxial deformation can be correlated with the commonly used domain of values, through the transformation $120^{\circ}-\gamma$.

Omitting the constant quasiparticle contribution, the relevant Hamiltonian for the system's dynamics can be written as \cite{Chir1}
\begin{equation}
\hat{H}_{chiral}=A_{1}\hat{I}_{1}^{2}+A_{2}\hat{I}_{2}^{2}+A_{3}\hat{I}_{3}^{2}-2A_{1}j\hat{I}_{1}-2A_{2}j'\hat{I}_{2}.
\label{Hal}
\end{equation}
This operator can be easily diagonalized in a basis of $2I+1$ rotation states $|IMK\rangle$, where $M$ and $K$ are total angular momentum projections in the laboratory and respectively body-fixed reference frames. The corresponding wave function is expanded in this basis as:
\begin{equation}
\Psi_{IMn}=\sum_{K=-I}^{I}\Lambda_{IKn}|IMK\rangle,
\end{equation}
where $n$ is the order and $\Lambda_{IKn}$ are the corresponding eigenvector components of a certain diagonalization solution.

In order to describe the dynamics associated with the constrained rotor Hamiltonian (\ref{Hal}), it is preferred to have a separated kinetic and potential energy as a function of a continuous variable. This can be achieved through a semiclassical approach detailed in Ref. \cite{Front}. Thus, applying a time dependent variational principle to the quantum Hamiltonian (\ref{Hal}), one obtains a classical energy function:
\begin{eqnarray}
&&\mathcal{H}_{I}(x,\varphi)=\frac{I}{2}(A_{1}+A_{2})+A_{3}I^{2}\\
&&+\frac{(2I-1)(I^{2}-x^{2})}{2I}(A_{1}\cos^{2}{\varphi}+A_{2}\sin^{2}{\varphi}-A_{3})\nonumber\\
&&-2A_{1}j\sqrt{I^{2}-x^{2}}\cos{\varphi}-2A_{2}j'\sqrt{I^{2}-x^{2}}\sin{\varphi}\nonumber,
\label{clase}
\end{eqnarray}
expressed in terms of a pair of canonically conjugated variables, $\varphi$ and $x$, which describe the direction of the total angular momentum vector in spherical coordinates. Therefore $\varphi$ is the azimuth angle $0\leq\varphi<2\pi$, while $x=I\cos{\theta}$ is the total angular momentum projection on the third intrinsic axis, defined by the corresponding polar angle $0\leq\theta<\pi$. The canonicity of the chosen variables is given by their Poisson bracket $\{\varphi,x\}=1$, identifying $x$ as a generalized momentum for the coordinate $\varphi$.

The classical energy function has a critical point at $x_{0}=0$. It is a minimum and therefore a stationary point for angular momentum values up to the critical spin:
\begin{equation}
I_{c}=\frac{\sqrt{A_{1}^{2}j^{2}(A_{2}-A_{3})^{2}+A_{2}^{2}j'^{2}(A_{1}-A_{3})^{2}}}{(A_{1}-A_{3})(A_{2}-A_{3})}+\frac{1}{2}.
\end{equation}
The dynamical phase with $I<I_{c}$ is called planar \cite{FrauMod}, because the average direction of the total angular momentum is restricted to the principal plane of the quasiparticle alignments. The next aplanar phase beyond $I_{c}$, splits the single stationary point into two chiral energy minima. The azimuth coordinate $\varphi_{0}$ of the stationary point in the planar phase is in general spin-dependent. It is determined as a solution of the following fourth order equation
\begin{align}
\left(\frac{\partial{\mathcal{H}_{I}}}{\partial{\varphi}}\right)_{0,\varphi_{0}}&=\frac{I(2I-1)}{2}(A_{2}-A_{1})\sin{2\varphi_{0}}\nonumber\\
&+2A_{1}Ij\sin{\varphi_{0}}-2A_{2}Ij'\cos{\varphi_{0}}=0\label{sol}.
\end{align}

\begin{figure}[th!]
\begin{center}
\includegraphics[width=0.45\textwidth]{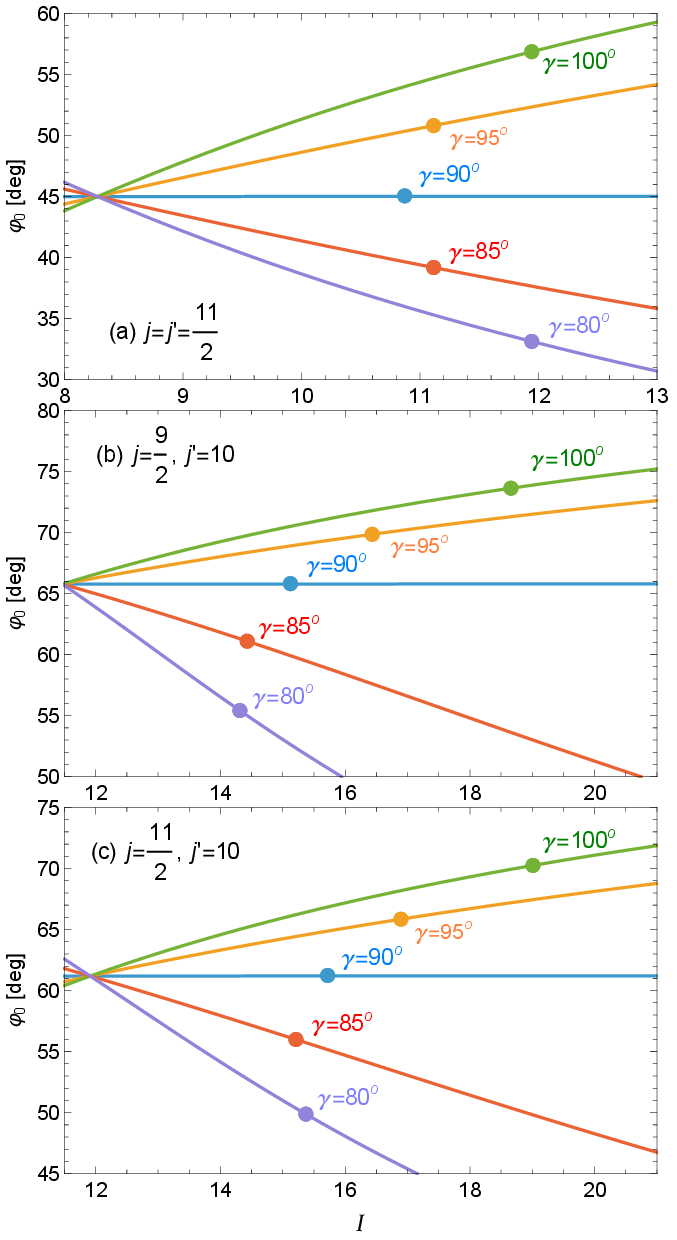}
\vspace{-0.7cm}
\end{center}
\caption{The solution of Eq. (\ref{sol}) as a function of total angular momentum for $j=j'=11/2$ (a), $j=9/2$, $j'=10$ (b), $j=11/2$, $j'=10$ (c) and $\gamma=80^{\circ},85^{\circ},90^{\circ},95^{\circ},100^{\circ}$ values of the triaxial deformation. The points denote the critical angular momenta up to which the coordinates $\varphi_{0}$ and $x_{0}=0$ correspond to a minimum for the classical energy function.}
\label{phi0}
\end{figure}

The dependence of the corresponding solution with total spin is presented in Fig. \ref{phi0} for various alignment and deformation configurations. When $\gamma=90^{\circ}$, one has $A_{1}=A_{2}=A_{\perp}$ and the azimuth coordinate of the planar stationary point acquires a constant value which can be analytically expressed as $\tan{\varphi_{0}}=j'/j$. For arbitrary $\gamma$ but equal quasiparticle alignments $j=j'$, the angle varies symmetrically for the prolate ($\gamma>90^{\circ}$) and oblate ($\gamma<90^{\circ}$) branches. Otherwise, its variation with $\gamma$ is weaker in the prolate branch which favors the larger quasiparticle spin.

\begin{figure*}[th!]
\begin{center}
\includegraphics[width=0.97\textwidth]{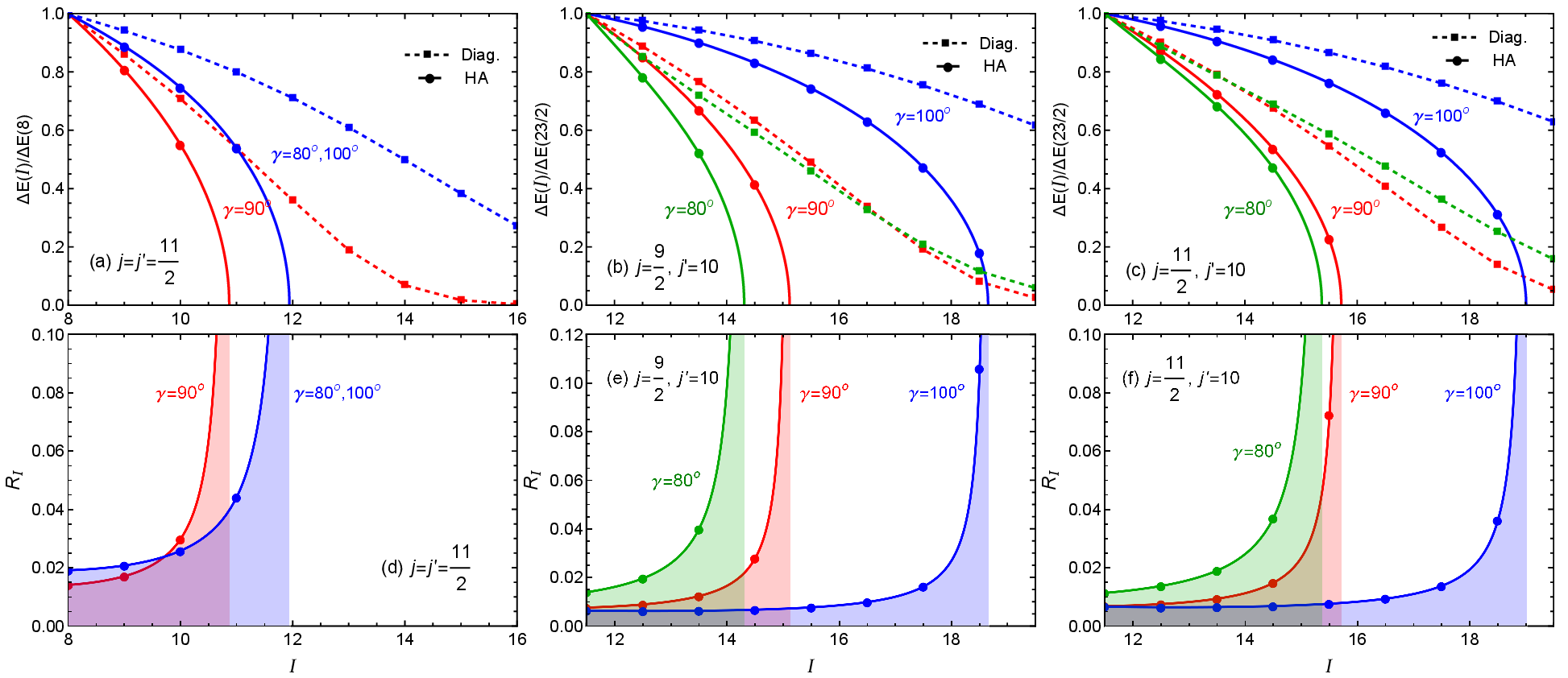}
\vspace{-0.7cm}
\end{center}
\caption{ (a-c) Energy splitting of the chiral partner bands determined from the exact diagonalization of the constrained rotor Hamiltonian and its harmonic approximation $\Delta E(I)=\omega_{I}$, for different values of triaxial deformation. (d-f) The ratio $R_{I}$ between the coefficients of the fourth and second order terms from the power expansion of the classical energy function in respect to $x$. The considered quasiparticle spins are $j=j'=11/2$ (a,d), $j=9/2,\,j'=10$ (b,e), and $j=11/2,\,j'=10$ (c,f).}
\label{Ew}
\end{figure*}

The classical trajectories of the total angular momentum in the space of the canonical coordinates are obtained from the integration of the equations of motion. In the planar phase, these trajectories encircle the single stationary point, which translate into a precession-like motion of the total angular momentum. Note that the equations of motion produce a variable velocity along the trajectory \cite{Coh}. The harmonic character of this time evolution was for example investigated in Refs. \cite{har1,har2}, using the tilted axis cranking in the framework of the covariant density functional theory. Using the analytical property of the present model, one will discuss here the harmonic condition in an algebraic manner. Having a single stationary point in the planar phase, one can perform a harmonic approximation of the classical energy function around this value:
\begin{eqnarray}
\tilde{\mathcal{H}_{I}}(x,\varphi)&=&\mathcal{H}_{I}(0,\varphi_{0})+\frac{1}{2}\left(\frac{\partial^{2}{\mathcal{H}_{I}}}{\partial{x}^{2}}\right)_{0,\varphi_{0}}x^{2}\nonumber\\
&&+\frac{1}{2}\left(\frac{\partial^{2}{\mathcal{H}_{I}}}{\partial{\varphi}^{2}}\right)_{0,\varphi_{0}}\left(\varphi-\varphi_{0}\right)^{2}.
\label{Haprox}
\end{eqnarray}
Given the canonical conjugate relationship between $x$ and $\varphi$, the above equation can be quantized into a Harmonic Oscillator problem. One can thus identify an effective mass and force constant as
\begin{equation}
\frac{s}{m_{I}}=\left(\frac{\partial^{2}{\mathcal{H}_{I}}}{\partial{x}^{2}}\right)_{0,\varphi_{0}},\,\frac{m_{I}\omega_{I}^{2}}{s}=\left(\frac{\partial^{2}{\mathcal{H}_{I}}}{\partial{\varphi}^{2}}\right)_{0,\varphi_{0}},
\end{equation}
where $s$ is a length implying the scalable nature of the harmonic oscillator problem and the uncertainty in the definition of the effective mass. The frequency of the obtained oscillator is independent of the scale
\begin{equation}
\omega_{I}=\sqrt{\left(\frac{\partial^{2}{\mathcal{H}_{I}}}{\partial{x}^{2}}\right)_{0,\varphi_{0}}\left(\frac{\partial^{2}{\mathcal{H}_{I}}}{\partial{\varphi}^{2}}\right)_{0,\varphi_{0}}}.
\label{frec}
\end{equation}
In this harmonic approximation (HA), the energy splitting between chiral partner states is given just by the above frequency, $\Delta E(I)=\omega_{I}$. The evolution with spin of the chiral frequency is depicted in Fig. \ref{Ew} (a-c), where it is also compared with the energy splitting deduced from the exact diagonalization of the rotor Hamiltonian with alignment constraints (\ref{Hal}). The reference moment of inertia $\mathcal{J}_{0}$ is eliminated by normalization to the energy of the first allowed angular momentum state. As a consequence of this comparison, the performance of HA is better for larger quasiparticle alignments and triaxial deformation closer to the axial symmetry, favoring the body fixed axis with the largest quasiparticle alignment. Larger quasiparticle alignments are achieved through the alignment of a quasiparticle pair from the proton or neutron $h_{11/2}$ particle orbital. The distribution of the two quasiparticles in this orbital is strongly localized around 10$\hbar$ angular momentum units \cite{Hammermesh}. The quality of the HA can be better seen from the quantity
\begin{equation}
R_{I}=\frac{1}{12}\left(\frac{\partial^{4}{\mathcal{H}_{I}}}{\partial{x}^{4}}\right)_{0,\varphi_{0}}\left/\left(\frac{\partial^{2}{\mathcal{H}_{I}}}{\partial{x^{2}}}\right)_{0,\varphi_{0}}\right.,
\label{rat}
\end{equation}
which relates the symmetric factors of the fourth and second order terms in the power expansion of the classical energy function in respect to the variable $x$. The conclusions regarding the energy splitting are mirrored in the dependence of this measure on angular momentum shown in Fig. \ref{Ew} (d-f) for different alignment and triaxiality configurations. It retains a slow linear increase at low spins which evolves into an exponential increase closer to the critical angular momentum $I_{c}$, where HA collapses. The same ratio for the azimuth angle $\varphi$ has a very weak variation with angular momentum in the range of few percents. This aspect is for example used in the approach of Refs. \cite{Chir1,Front,Chir2,Chir3}, where the classical function is harmonically approximated only for the $\varphi$ coordinate.

The special case of $\gamma=90^{\circ}$ provides also a closed expression for the associated chiral frequency:
\begin{equation}
\omega_{I}(\gamma=90^{\circ})=\sqrt{2A_{\perp}\tilde{j}\left[2A_{\perp}\tilde{j}-(2I-1)(A_{\perp}-A_{3})\right]},
\end{equation}
where $\tilde{j}=\sqrt{j^{2}+j'^{2}}$ is the resultant quasiparticle spin. In this expression one can readily recognize the wobbling frequency from Refs. \cite{Frau1,Wob1}. As a matter of fact, the similarity between the transverse wobbling and the chiral vibration phase was mentioned previously in a series of studies \cite{Chir2,col1,col2,Raduta1}.

Choosing a representation where projection $x$ is the active variable, one can write the corresponding wave function in the momentum space as:
\begin{equation}
\phi_{In}(x)=\frac{1}{\sqrt{2^{n}n!}}\sqrt{\frac{\alpha_{I}}{\sqrt{\pi}}}e^{-\frac{\alpha_{I}^{2}x^{2}}{2}}H_{n}(\alpha_{I} x),\label{howf}
\end{equation}
where $H_{n}$ are Hermite polynomials, while
\begin{equation}
\alpha_{I}=\frac{1}{\sqrt{m_{I}\omega_{I}}}=\sqrt{s}\left[\left(\frac{\partial^{2}{\mathcal{H}_{I}}}{\partial{x}^{2}}\right)_{0,\varphi_{0}}\left/\left(\frac{\partial^{2}{\mathcal{H}_{I}}}{\partial{\varphi}^{2}}\right)_{0,\varphi_{0}}\right.\right]^{1/4}.
\end{equation}
The quantum numbers $n=0$ and $n=1$ identify the yrast state and respectively its chiral partner of the same angular momentum $I$. The oscillator wave function (\ref{howf}) defines a density of probability
\begin{equation}
\rho_{In}(x)=\left|\phi_{In}(x)\right|^{2},
\end{equation}
which can be used to generate the coefficients
\begin{equation}
\tilde{\Lambda}_{IKn}=\phi_{In}(K)\left[\sum_{K=-I}^{I}\rho_{In}(K)\right]^{-1},
\end{equation}
of the new total wave function expansion in rotation matrices:
\begin{equation}
\tilde{\Psi}_{IMn}=\sum_{K=-I}^{I}\tilde{\Lambda}_{IKn}|IMK\rangle.\label{Expand}
\end{equation}

\begin{figure}[th!]
\begin{center}
\includegraphics[width=0.47\textwidth]{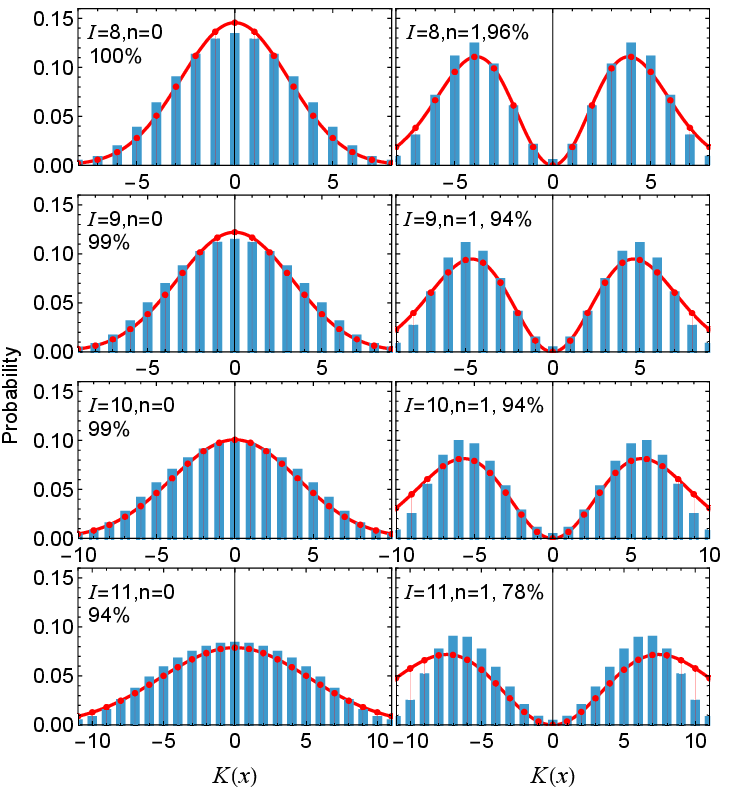}
\vspace{-0.5cm}
\end{center}
\caption{The total wave function components $|\Lambda_{IKn}|^{2}$ and $|\tilde{\Lambda}_{IKn}|^{2}$ spread over $K=-I,-I+1,...,I-1,I$, determined from the exact diagonalization of equation (\ref{Hal}) (blue histogram) and deduced from the HA (red circles and continuous curve) when $j=j'=11/2$, $\gamma=100^{\circ}$ and the scale is fixed as $s=0.02$. First column shows the ground states with $n=0$, while the second column shows the excited states with $n=1$. The listed percentage represents the fraction of the HA probability $\rho_{In}(x)$ encompassed in the physical $[-I,I]$ interval.}
\label{Prob}
\end{figure}

Note that the probability distribution is renormalized in the definition of the expansion coefficients. This is due to the fact that although the expansion is made in the limited interval $x\in[-I,I]$, the oscillator wave function is normalized on the whole space $(-\infty,\infty)$. This is where intervenes the scale $s$. Its role is to scale the oscillator wave function to the physically relevant interval. This is done by setting the value of $s$ to the minimum which assures that the whole probability distribution for the classical minimal angular momentum value $\tilde{j}=\sqrt{j^{2}+j'^{2}}$ is completely contained in the $[-\tilde{j},\tilde{j}]$ bounds. Fig. \ref{Prob} shows that such a scheme makes the probability distribution of the adopted harmonic approximation very similar to the results obtained from the diagonalization of the constrained rotor Hamiltonian (\ref{Hal}). The HA is obviously better for low spin states in consistence with the comparison made in Fig. \ref{Ew}, but also for the yrast states $n=0$. This HA picture offers the possibility to interpret the quantum states as quantized harmonic oscillations of the total angular momentum vector. Due to the discrete nature of the angular momentum projection, the original quantum picture cannot provide such a phenomenological insight.

\begin{figure*}[th!]
\begin{center}
\includegraphics[width=0.97\textwidth]{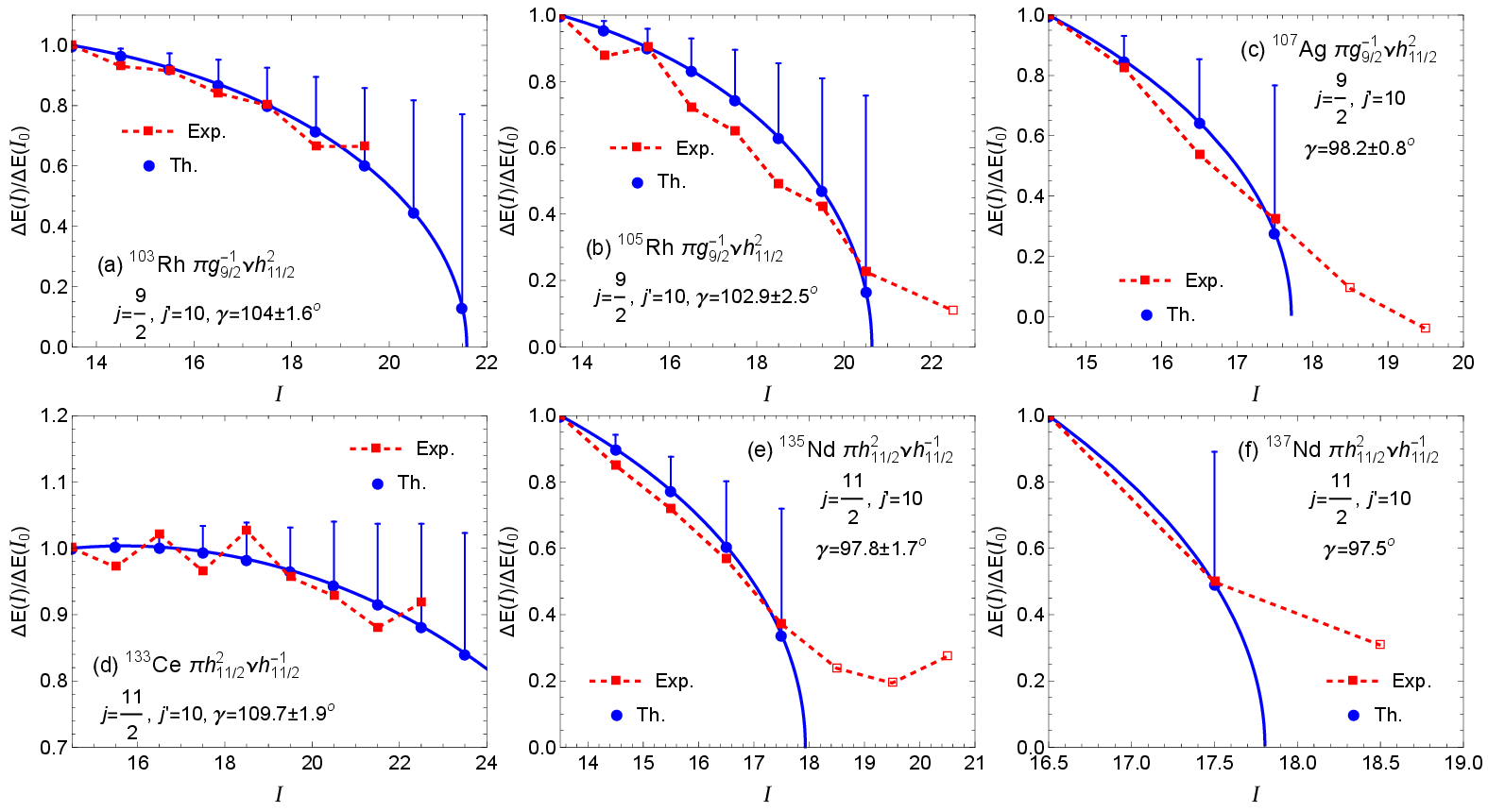}
\vspace{-0.5cm}
\end{center}
\caption{The angular momentum dependence of the experimental energy splitting $\Delta E(I)$ between the chiral partner bands of the $^{103,105}$Rh \cite{103Rh,105Rh}, $^{107}$Ag \cite{107Ag}, $^{133}$Ce \cite{133Ce}, and $^{135,137}$Nd \cite{135Nd,137Nd} nuclei, is compared with the values calculated within the HA. Both quantities are normalized to the data point corresponding to the lowest considered energy splitting of angular momentum $I_{0}$. The fitted triaxiality measure is shown for each nucleus with its statistical error. Open symbols denote data not considered in the fitting procedure, while the vertical bars show the departure from the exact diagonalization results.}
\label{ExpW}
\end{figure*}

Given the above spectral analysis, the experimental realization of the harmonic chiral vibration is sought between $A\sim100$ and $A\sim130$ nuclei \cite{Xiong}, exhibiting chiral partner bands built on large quasiparticle alignments. The central criterium is represented by an energy splitting between chiral states which decreases with spin. The $\pi g_{11/2}^{-1}\nu h_{11/2}^{2}$ chiral bands from the $^{103,105}$Rh \cite{103Rh,105Rh} and $^{107}$Ag \cite{107Ag} nuclei and those built on the $\pi h_{11/2}^{2}\nu h_{11/2}^{-1}$ quasiparticle configuration from the $^{133}$Ce \cite{133Ce} and $^{135,137}$Nd \cite{135Nd,137Nd} nuclei came out of this selection. For numerical applications, it is important to establish the range of HA applicability. This is done by eliminating the data points close to the $\Delta E=0$ and those exhibiting an obvious change from the monotonic decrease with spin of the energy split. The selected states are then fitted with the chiral frequency (\ref{frec}). The theoretical chiral frequency depends only on the triaxiality measure $\gamma$ and the empirical factor $\mathcal{J}_{0}$ which is eliminated from the fitting procedure by a suitable normalization of the considered energy differences.

The numerical results are visualized in Fig. \ref{ExpW} in comparison with the corresponding experimental data regarding the energy split $\Delta E$. The obtained $\gamma$ parameter suggests a high to moderate triaxiality, favoring the prolate branch. The $120^{\circ}-\gamma$ values resulted for the Rh isotopes are consistent with those adopted in Refs. \cite{Qi,Kuti}. The triaxiality used in PRM calculations of Ref. \cite{MultiAg} for the $^{107}$Ag nucleus is very similar to the presently fitted value, although other previous theoretical studies suggest a very low triaxial deformation \cite{Jere,Ma}. The lowest $\gamma$ value and respectively the highest triaxiality is reported for the $^{137}$Nd nucleus which is however presented only as a complementary to the case of $^{135}$Nd with more considered data points. The fact that both Nd nuclei are found to be closest to the maximal triaxiality is supported by previous studies \cite{135Nd,137Nd}. Similarly, the lowest $120^{\circ}-\gamma$ value associated with chiral bands of the $^{133}$Ce is consistent with the deformation determined from the Relativistic Mean Field calculations of Ref. \cite{Agn} and the older description of these bands as generated by purely prolate configurations \cite{Ma2}. This aspect allows for the description of extended sets of $^{133}$Ce partner states which are far from degeneracy. It is then expected that the present harmonic chiral vibration model could provide useful extrapolation to higher spin states.

It must be mentioned that the simple HA formalism predicts that chiral bands of multiple vibration phonons $n>1$ are equidistant in energy, differing in units of $\omega_{I}$. The possible existence of such bands is for example supported by the experimentally suggested two phonon transverse wobbling bands \cite{163Lu,165Lu,135Pr,Kr}.

\begin{figure}[th!]
\begin{center}
\includegraphics[width=0.48\textwidth]{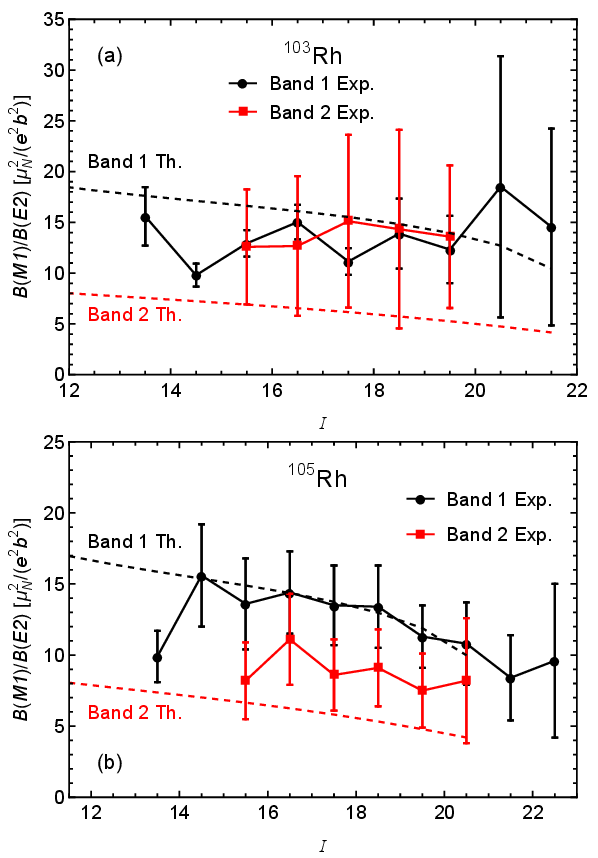}
\vspace{-1.1cm}
\end{center}
\caption{Comparison between experimental and theoretical $B(M1)/B(E2)$ ratio within the yrast (1) and excited (2) bands of $^{103}$Rh (a) \cite{103Rh} and $^{105}$Rh (b) \cite{Xiong}. The employed empirical quadrupole moments are $Q_{0}=7.7\,e\,b$ and respectively $7.6\,e\,b$, while the triaxial parameter $\gamma$ is taken from Fig. \ref{ExpW}.}
\label{Tranz}
\end{figure}

\begin{figure}[th!]
\begin{center}
\includegraphics[width=0.48\textwidth]{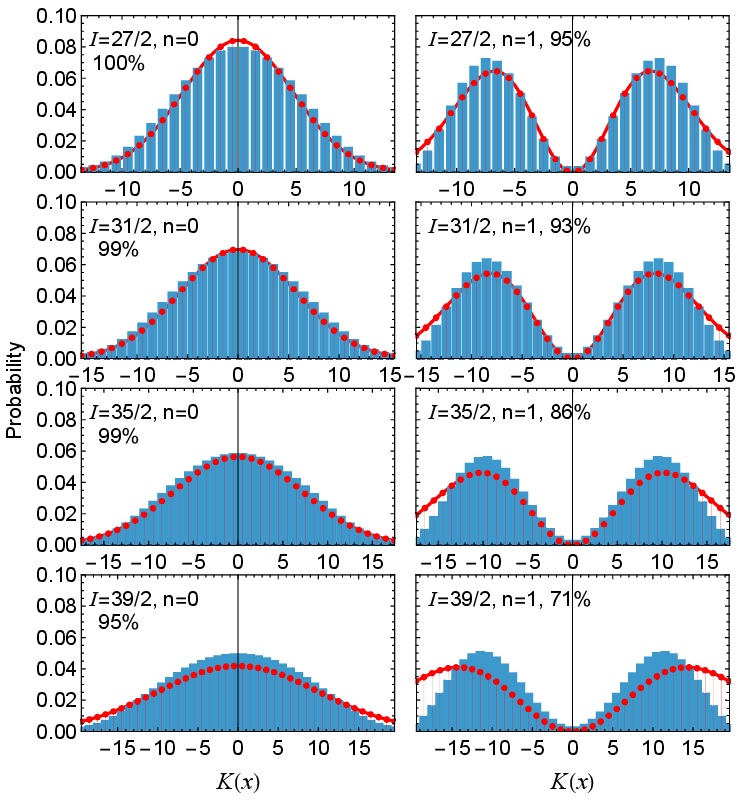}
\vspace{-1.1cm}
\end{center}
\caption{Same as in Fig. \ref{Prob}, but for a selection of states within the $j=9/2$, $j'=10$ and $\gamma=102.9^{\circ}$ geometry corresponding to the numerical application for the $^{105}$Rh nucleus. The same scale $s=0.02$ is used.}
\label{Rh}
\end{figure}

The triaxial parameter $\gamma$ extracted from the experimental energy levels is further used to define the coefficients of the total wave function expansion (\ref{Expand}). Having the rotation matrix distribution, one can now calculate the $E2$ and $M1$ electromagnetic transitions using the following operators
\begin{align}
T_{1\mu}(M1)&=\sqrt{\frac{3}{4\pi}}\mu_{N}\sum_{\nu=0,\pm1}\left(g_{eff}j_{\nu}+g'_{eff}j'_{\nu}\right)D^{1}_{\mu\nu},\\
T_{2\mu}(E2)&=Q_{0}\left[\cos{\gamma}D_{\mu0}^{2}+\frac{\sin{\gamma}}{2}\left(D_{\mu2}^{2}+D_{\mu-2}^{2}\right)\right].
\label{ope2}
\end{align}
$g_{eff}=g_{j}-g_{R}$ and $g'_{eff}=g_{j'}-g_{R}$ are effective gyromagnetic factors gathering the corresponding values of the core and of the single-particle orbitals, $\mu_{N}$ is the nuclear magneton, $D_{\mu\nu}^{1(2)}$ are Wigner rotation matrices, while $Q_{0}$ denotes an empirical quadrupole moment. The gyromagnetic factor of the core is determined as the ratio between the corresponding charge and mass numbers, while its quasiparticle counterpart is calculated by adopting a 0.6 quenching factor for the free spin gyromagnetic factors.

The ratio of in-band magnetic and electric transition probabilities from the same state is commonly used to describe the electromagnetic properties of the chiral partner bands. Its evolution with spin is exemplified in Fig. \ref{Tranz} for the $^{103,105}$Rh nuclei. The calculations are performed using the $s=0.2$ scaling factor. For a meaningful comparison with experimental data, the empirical factor $Q_{0}$ is fitted for the best agreement. The mostly linear decreasing dependence with spin of the $B(M1)/B(E2)$ ratio retains an almost constant difference between the values of the two bands, with the highest values associated with the yrast band. The theory overestimates this difference for the $^{105}$Rh, but reproduces very well its evolution with spin. The theory predicts a similar behaviour for the $^{103}$Rh nucleus, which is not supported by data but still fall in the most of the experimental error bars. The values of the two bands approach each other in the vicinity of the critical angular momentum marking the termination of the chiral vibration mode. This is consistent with the experimental data for the $^{105}$Rh nucleus, suggesting thus the gradual establishing of static chirality. Analyzing Fig. \ref{Rh}, one can see that the termination of the planar phase is characterized by a flattening of the probability distribution of the ground state and an increase of the weights corresponding to the $x(K)\rightarrow I$ components of the excited state. The later feature is exaggerated by the HA in respect to the exact diagonalization results, especially for states closer to $I_{c}$. Although Figs. \ref{Prob} and \ref{Rh}, look similar, one must note that the set of considered states is significantly larger in the case of the $^{105}$Rh. It must be mentioned that there are available experimental data on transitions, also for both bands of $^{133}$Ce and $^{135}$Nd nuclei. However, the measured transition ratios in these nuclei are higher in the excited band, a fact which is in contradiction with the predictions of the present model.

In conclusion, we showed that the low lying states from chiral partner bands with a sizable energy shift between them, can be described by means of a harmonic oscillation of the total angular momentum vector. The proposed simple picture is more appropriate for chiral bands built on large quasiparticle alignments and moderate triaxiality. These conditions assure that the chiral vibration phase, with a planar equilibrium direction of the total angular momentum, is extended over a larger set of states. The analytical properties of the adopted harmonic approximation provide strong spectroscopic signatures. More precisely, the energy difference between chiral partner states has a parabolic decrease with spin which can be used to ascertain the triaxial deformation. On top of this, there is a constant difference in the $B(M1)/B(E2)$ ratio of in-band transitions, in favor of the yrast band. Numerical applications identified the $^{105}$Rh nucleus as being the best example of a harmonic chiral vibrator.

\section*{Acknowledgments}
This work was supported by grants of the Ministry of Research, Innovation and Digitization, CNCS - UEFISCDI, project number PN-IV-P1-PCE-2023-0273, within PNCDI IV, and project number PN-23-21-01-01/2023.

\end{document}